\documentclass[12pt,reqno]{amsart}
\usepackage{amsmath,amsthm,amsfonts}
\usepackage{txfonts}
\usepackage[T1]{fontenc}
\usepackage[utf8]{inputenc}
\usepackage{a4wide}
\usepackage{amssymb}
\usepackage{caption}
\usepackage{epsfig}
\usepackage{subfigure}
\usepackage{graphics,color,psfrag}
\usepackage{float}
\usepackage[all]{xy}
\usepackage{booktabs,longtable}
\usepackage{epstopdf}
\usepackage{hyperref}
\usepackage[hyphenbreaks]{breakurl}

\allowdisplaybreaks[4]
\newcommand{\ie}{\textit{i.e.~}}

\newcommand{\LL}{\mathcal{L}}

\newcommand{\ds}{\displaystyle}

\theoremstyle{definition}

\theoremstyle{remark}

\newtheoremstyle{bgtheorem}{3pt}{3pt}{\itshape}{}{\bfseries}{}{ }%
{\thmname{#1}\thmnumber{ #2}\thmnote{ (#3)}}
\theoremstyle{bgtheorem}

\newtheoremstyle{case}{5pt}{5pt}{}{}{\scshape}{ }{ }%
{\thmnote{[#3]}}
\theoremstyle{case}

\numberwithin{equation}{section}

\begin{document}

\title[Spread of mobile viruses]{
A Monte Carlo method for the Spread of Mobile Malware
}

\author{Alberto Berretti}
\address{Dipartimento di Ingegneria Civile ed Ingegneria Informatica\\
Universit\`{a} di Tor Vergata, Roma, Italy}
\email{{\tt berretti@disp.uniroma2.it}}

\author{Simone Ciccarone}
\address{Dipartimento di Ingegneria Civile ed Ingegneria Informatica\\
Universit\`{a} di Tor Vergata, Roma, Italy}
\email{{\tt simone.ciccarone@uniroma2.it}}

\begin{abstract}
A new model for the spread of mobile malware based on proximity (\ie Bluetooth,
ad-hoc WiFi or NFC) is introduced. The spread of malware is analyzed using a 
Monte Carlo
method and the results of the simulation are compared with those from 
mean field theory.
\end{abstract}

\maketitle

\thispagestyle{empty}

\vspace{0.5cm}

\section{Introduction}\label{sect:intro}

Mobile computing platforms (from simple featurephones to smartphones to
tablets) are becoming ubiquitous and ever more capable, and they are slowly eroding 
the predominance of the personal computer, especially at the notebook level. As they are 
becoming more capable and more common, they become the target of malware (viruses, worms,
etc.). In fact, you can't infect a device which is not capable of executing arbitrary binary
programs, which renders old fashioned, dumb, cellular phones relatively 
safe, and moreover there is no incentive in looking for exploits and developing 
malware for 
platforms with a limited market; but the always increasing raw computing power 
of the modern smartphones (multi-core processors, gigabytes of RAM, and recently 
even 64 bit processors) make them as powerful as yesterday's personal computers, 
and their diffusion makes them a valuable target, one worth exploiting
by organized crime. 

Mobile malware is not new. Years ago, Symbian was the most popular smartphone 
operating system and several viruses have been developed to infect 
Symbian-based phones. It is believed that Cabir\cite{FS}, 
developed by an unknown hacking group around 2004, was the first such virus. It infected Symbian 
Series 60 phones with Bluetooth enabled, and it spread to nearby phones but 
required (usually, but not in all brands and models of phones) user intervention 
to accept the download of the executable over Bluetooth. Cabir didn't do 
anything but spreading, eventually rendering the UI of the phone useless for the 
continuous requests to accept the download of the executable or draining 
the battery. In 2005 a similar virus dubbed ``CommWarrior''\cite{Sym}, always 
targeting the Symbian Series 60 platform, spread using both Bluetooth (infecting 
arbitrary nearby phones) and MMS (following therefore the social graph of the 
user by choosing randomly contacts from the user's phonebook). 
Several more viruses have been developed since, even if platforms, infection 
mechanisms and purposes have changed to keep up with the latest trends in mobile 
computing. 

Malware which explicitly targets mobility can exploit two characteristics: 
follow the social graph of the user exploiting access to the user's phonebook 
entries (``social malware'', sometime also called ``topological malware'' to 
stress the influence on the topology of the social graph on its spread), or use 
protocols based on proximity like Bluetooth, ad-hoc WiFi connections or NFC 
(``proximity-based malware''). In this paper we focus on this last aspect of 
mobile malware, developing a ``microscopic'', stochastic model for the 
propagation of the infection suitable for Monte Carlo simulations.

\section{Definition of the model}\label{sect:def}

When modeling infections spread via proximity it is natural to consider 
percolation models, where susceptible objects occupy sites on a regular lattice 
-- or are distributed with different topologies: for example a graph -- and the 
infection spreads from an infected site to neighbour sites. This approach 
doesn't take into account mobility, where neighbours change as each susceptible 
object moves around. 

A simple and realistic way to take into account mobility is to make each 
object perform a random walk in a regular lattice. As objects get to approach, 
an infected one can pass on the infection to each of its -- temporary -- 
neighbour, with a given probability. 

To be definite, we assume that $N$ susceptible objects perform a random walk on 
a square portion of a two-dimensional lattice $\LL = 
\{0,\ldots,L-1\}\times\{0,\ldots,L-1\}$. The density of objects is therefore $d 
= N/L^2$. These objects are initially placed in some arbitrary way on the 
lattice, for example are uniformly distributed. Time is discrete, 
and each object performs a random walk moving with equal probability in one of 
the four possible directions into one of the four nearest neighbour sites. 

We have to deal somehow with the finite size of the box $\LL$ in which the 
objects move. ``Free'' boundary conditions, in which each object is free to 
leave the box $\LL$, would deplete the box itself with probability one after a 
finite amount of time, so one would have to take into account the appearance of 
new objects which \emph{move into} $\LL$: this would give rise to a model with 
variable number of objects, something like a ``grand canonical ensemble'' in 
statistical mechanics. We prefer for the moment to avoid the complexity of 
dealing with disappearing and reappearing new objects, which we consider 
inessential to the problem, so we use periodic boundary conditions: as one 
object moves out of $\LL$ on one side, it reappears from the opposite side of 
the box; the random walk happens therefore on a torus. Another possibility 
would be to have objects bouncing when they reach the boundary: while this is 
relatively simple to take into account in the simulation, again it would add 
complexity to the model without really changing in any significant way the 
phenomenology of the model (as we tested).

Each object can be in one of two states: healthy or infected. As objects move 
into the same site the infection can spread from one of the infected object to 
each one of the object which occupy the same site with a given probability $p$. 
Of course, more than two objects can be on the same site at a given time: in 
this case we consider separately all pairs as a possible source of infection 
(\ie if we have three objects on a site, two infected and one healthy, then we 
test for a possible infection of the healthy object \emph{twice} independently, 
because of the two possible sources of infection).

As all objects eventually intersect their trajectories, and eventually get 
the infection, all objects sooner or later would become infected. We therefore 
have to take into account also the chance that a given object heals itself 
(perhaps because the infection has been detected and dealt with). So at each 
(discrete) instant of time each infected object has a probability $q$ of 
healing. 

We therefore have three parameters which determine the spread of the infection 
on a given box $\LL$ (besides the size of the box): the density of objects $d$, 
the probability of infection $p$ and the probability of healing $q$. We look at 
the case in which the box is large, \ie what would be called an ``infinite 
volume limit'' or ``thermodinamical limit'' in statistical mechanics. This is 
basically a SIS model, using standard epidemiological terminology: recovered 
objects can get infected again and don't get to be immune, as in so-called SIR 
models. If, using our mobility model, we were to use a SIR model, all objects 
eventually would be infected, recover and never get infected again and the 
epidemics would stop with probability one in a finite amount of time, 
independently on the side of the box. 

This model is, mathematically, a Markov chain whit a huge state: if we have 
$M$ objects in our box, the different possible configurations are $2^M$. It is 
clearly possible to transition from any state to any state, with the exception 
of the state in which all objects are healthy and so there isn't anymore a way 
to get infected: this is an absorbing state that eventually the Markov chain 
will reach; for instance, if we have $N$ infected objects, with probability 
$q^N$ (extremely small but non-zero) they could get all healthy at the same 
time. We 
believe that the probability of reaching such a state in a given fixed amount 
of time, given some fixed values of $p$, $q>0$ and $d$, is exponentially small 
in the volume of the box and so negligible in the ``thermodinamical limit'' 
that we are considering. 

The choice of a square lattice and a simple random walk over it as a 
mobility model is somewhat arbitrary and motivated basically by mathematical 
simplicity. More complicated mobility models could be (and have been) devised. 
But as we are interested in \emph{qualitative} features, and not in exact 
\emph{quantitative} features of specific, realistic models, we concentrate 
our attention on a mathematically simple model which, while avoiding the 
complexities of a realistic one, keeps its qualitative features, much in the 
spirit of most statistical mechanical models commonly used in mathematical 
physics. As a byproduct, our simulation code is simpler, faster and more 
efficient. We also emphasize that our model, being based on a discrete random 
walk in a lattice, is a discrete time simulation. 

\section{Related work}\label{sect:relwork}

A few papers have studied simulations of SIS models via random walks. In 
\cite{BettstetterWagner} the spatial distribution of nodes in a random waypoint 
model is studied, and it is shown to be inhomogeneous. The authors only take 
into account the distribution of the agents, without ever considering 
propagation of infection. They use physical units of measures in the 
simulation, taking into account an area of $1000\times 1000$ square meters 
divided into square cells whose side is $20$ meters, so they actually use 
$50\times 50$ lattice (and so quite small). 

In \cite{VahdatBecker} the transmission of messages in a network of mobile 
nodes is studied. The authors again use physical units and so they consider an 
area of $1500\times 300$ meters where $50$ agents move using a random waypoint 
model, with a transmission radius which varies between 10 and 2500 meters (so 
again the effective size of the region is quite small). Using a simulation the 
authors study the mean time to deliver a message and the number of hops 
necessary to reach destination. 

In \cite{JaffryTreur} a SIR model with different mobility models is considered, 
with a populations of up to 1000 objects and a time of up to 1000 discrete 
units, and they look at the average number of immunized objects. 

In \cite{Valler} are used again physical units: they consider an area of 
dimension $200 \times 200$ m with an interaction radius of $5$ m, so the model 
can be compared to a lattice model with dimension $40 \times 40$, that is 
rather small. Several mobility models are taken into account, suitable for a 
continuous space and time model. The author computes an approximate formula 
for the infection threshold and shows that it depends only by the ratio of the 
probabilities of infection and disinfection. While similar in general 
conception, our model is rather different 
(and quite simpler to simulate: in fact we could handle a simulation with a 
much 
larger number of agents and area) as it is based on random walks on a 
lattice, and we found instead a more complex dependence 
of the epidemic threshold from infection and disinfection probabilities, even 
in the mean field theory approximation.

\section{The simulation}\label{sect:simul}

The code for the simulation has been written in C to achieve optimum 
performance. We ran in on a small cluster using at most eight 
computational cores (four Intel Xeon dual core processors at 3 GHz) and on a 
small personal workstation (with an Intel I3 processor at 3.1 GHz), all running 
Linux.

We used square lattices of sizes from 16x16 up to 512x512. The results from 
lattices of different size have been compared and we 
observed that they do not change significantly for 
lattices of size higher than 64 x 64, while they are more volatile for lattices 
of smaller sizes: so a kind of ``thermodinamical limit'' is practically 
achieved already for this size. So we settled for a size of 128x128. 

The population density was chosen between 0.1 and 1 in step of 0.1, and also a 
few run at higher densities have been performed (with densities equal to 2, 5 
and 10). Note that densities higher than 1 imply that 
most sites are occupied by more than one agent, which is entirely possible 
within our model. 

The built-in random number generator of the compiler has been used for the 
simulations. 

As expected, the limit fraction of infected agents $f_{\infty}$ doesn't depend 
on its initial value $f_0$, so we took $f_0=0.2$ in all production runs. We 
also took a uniform initial distribution of the agents, as we expect that at 
equilibrium the agents are uniformly distributed (but we can't say anything 
about eventual fluctuations). 

We performed an autocorrelation analysis of the data from each simulation, to 
compute its autocorrelation time $\tau$. This is of course what must be done in 
any dynamic Monte Carlo simulation to insure that data points are taken from an 
equilibrium distribution and that they are taken sufficiently far apart so that 
they can be considered independent. In our case, moreover, the time to reach 
the equilibrium is an interesting quantity \emph{per se}. 
To compute autocorrelation times, we used a Python version of the 
\texttt{acor} package written by Jonathan Goodman \cite{Goodman}. 

\section{Mean Field Theory}\label{sect:meanfield}

We can approximate our model using a kind of ``mean field theory'' approach, 
which is expected to have a \emph{qualitatively} agreement with the complete, 
exact model. In this approximation we consider \emph{a single object}, whose 
evolution is based on the average behaviour of the rest of the system.

In average, each objects undergoes $\approx d$ intersections at 
each time, with each intersection giving a chance to get infected if it 
happens with an infected object. As $f = N/M$ is the fraction of the 
infected objects, at each time each object \emph{approximately} intersects with 
an infected one ``$fd$ times''. 

Therefore approximately the probability that an object gets infected is $p' = 
1 - (1-p)^{fd}$ -- pretending that $fd$ is an integer --, that is $1$ minus the 
probability of \emph{never} getting 
infected in each of its $fd$ intersections with an infected object. 

So considering each object individually, let $X = H$ or $I$ denote the status 
of an object ($H$ for healthy and $I$ for infected). The following diagram 
explains the transition to the new state with each probability:

{\small \begin{equation*}
\xymatrix{
&&&&H&p'q\\
&&I\ar[rru]^{q}\ar[rrd]^{1-q}&&&\\
&&&&I&p'(1-q)\\
X\ar[rruu]^{p'}\ar[rrdd]^{1-p'}&&&&&\\
&&&&H&(1-p')q\\
&&X\ar[rru]^{q}\ar[rrd]^{1-q}&&&\\
&&&&X&(1-p')(1-q)
}
\end{equation*}}

The dynamics of a single object can therefore be approximated by a much simpler 
Markov chain, where the state space is just the set $\{H,I\}$ (being healthy 
or being infected) and the transition probabilities are given by:

\begin{table}[H]
\centering
\begin{tabular}{ccc}
State at $t$&State at $t+1$&Probability\\
\midrule
$H$&$H$&$p'q+(1-p')q+(1-p')(1-q)=1-p'+p'q$\\
$H$&$I$&$p'(1-q)=p'-p'q$\\
$I$&$H$&$p'q+(1-p')q = q$\\
$I$&$I$&$p'(1-q)+(1-p')(1-q)=1-q$\\
\end{tabular}
\end{table}

The transition matrix is therefore:
\[
M = \begin{bmatrix}
    1-p'+p'q & p'-p'q \\
    q & 1-q
    \end{bmatrix}.
\]

This is an ergodic Markov chain whose invariant probability distribution is 
given by the normalized eigenvector of the eigenvalue $1$, given by:
\[
\begin{pmatrix}\ds\frac{q}{p'+q-p'q} \\ 
\ds\frac{p'-p'q}{p'+q-p'q}\end{pmatrix}.
\]
This Markov chain therefore approach an equilibrium state with a probability of 
having an infected object equal to $\ds\frac{p'-p'q}{p'+q-p'q}$. We take this 
value as the mean-field approximation for the fraction of the infected objects:
\[
f_{\text{MF}} = \frac{p'-p'q}{p'+q-p'q}. 
\]
As $p'$ depends on the fraction of the infected objects itself, after a few 
elementary steps we obtain a transcendental equation for $f_{\text{MF}}$:
\begin{equation}\label{fmf}
f_{\text{MF}} = 1 - \frac{q}{1-(1-q)(1-p)^{f_{\text{MF}}d}}. 
\end{equation}

Please note that besides assuming a perfect uniform distribution of agents and 
also a perfect uniform distribution of \emph{infected} agents, we also assumed 
$fd$ to be an integer, which of course is another approximation. Moreover in 
mean field theory there is always a chance of getting infected (the Markov 
chain is
actually really ergodic). 

To compute the epidemic threshold in mean field theory, we start by observing
that equation \eqref{fmf} always has a solution $f=0$. So we are in the 
\emph{epidemic regime} it there is \emph{another} solution 
$f = f_{\text{MF}} > 0$, for given values of $p$, $q$ and $d$.

To study the existence of solutions to \eqref{fmf}
we consider the 
intersection of the graph of:
\[
\phi(f) = 1 - \frac{q}{1-(1-q)(1-p)^{fd}}
\]
with the bisectrix of the first quadrant 
with $0\leq f\leq 1$.
 
By trivial calculations, we have that for any physical values of $p$, $q$ and $d$ 
($0\leq p\leq 1$, $0\leq q\leq 1$, $d \geq 0$) $\phi'(f)>0$ and $\phi''(f)<0$ and 
of course $\phi(0)=0$, $\phi(f)<1$. Therefore if $\phi'(0)>1$ we have another solution
$f = f_{\text{MF}} \in (0,1)$, while if $\phi'(0)\leq 1$ the only solution to 
\eqref{fmf} is $f=0$. So the condition $\phi'(0)=1$ determines the epidemic 
threshold in mean field; a trivial calculation gives:
\[
q_0 = \frac{d\log\frac{1}{1-p}}{1+d\log\frac{1}{1-p}},
\]
with the epidemic thriving if $q<q_0$ and extinguishing if $q>q_0$. 

\begin{figure}[ht]
        \centering
  \subfigure[$d=0.3$.]{
    	\psfig{file=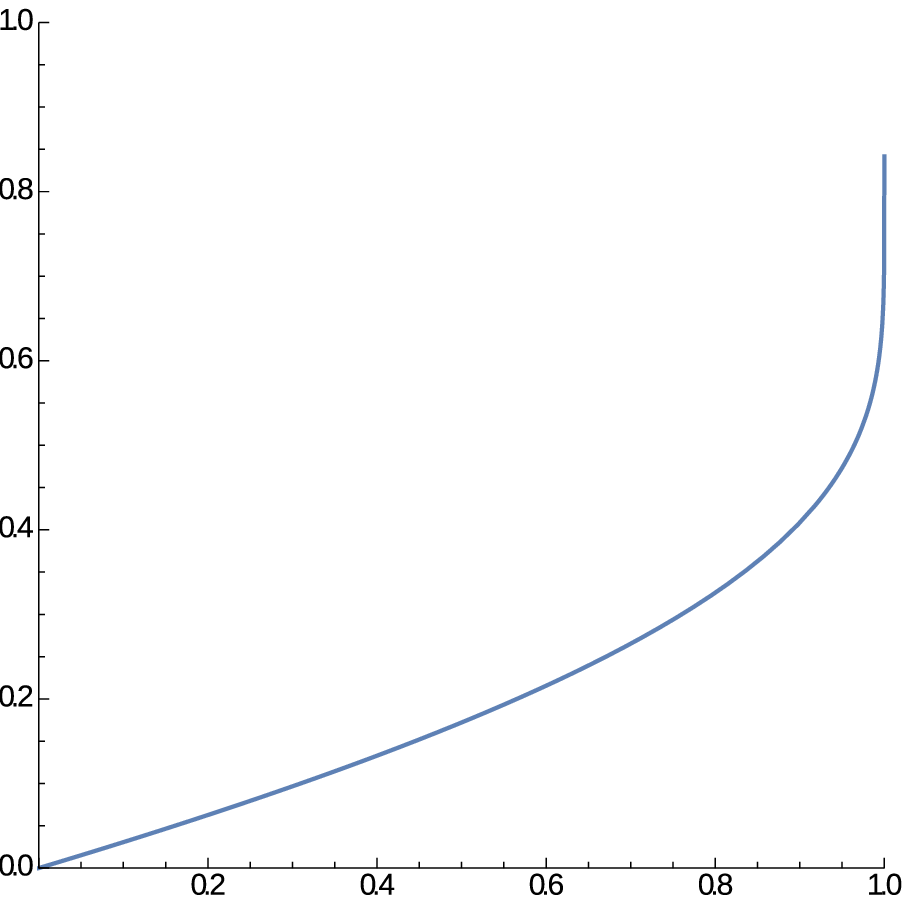,height=1.95in}}
     \hspace{0.3 in}
  \subfigure[$d=0.5$.]{
    	\psfig{file=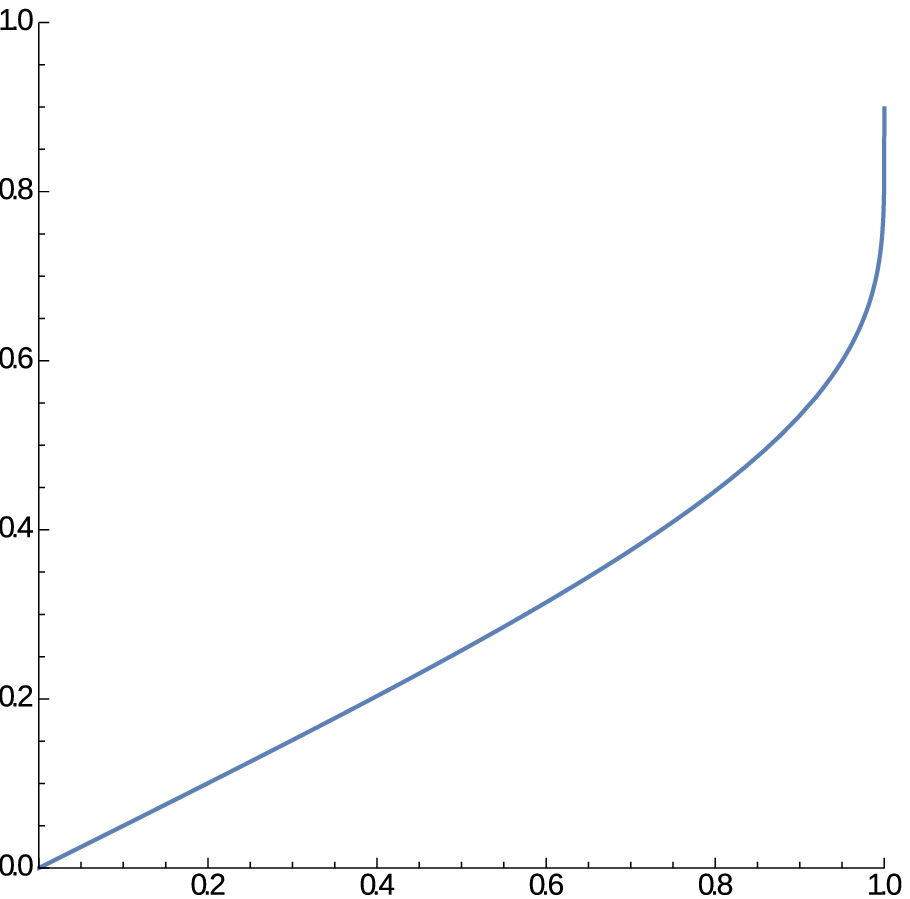,height=1.95in}}
  \subfigure[$d=0.7$.]{
    	\psfig{file=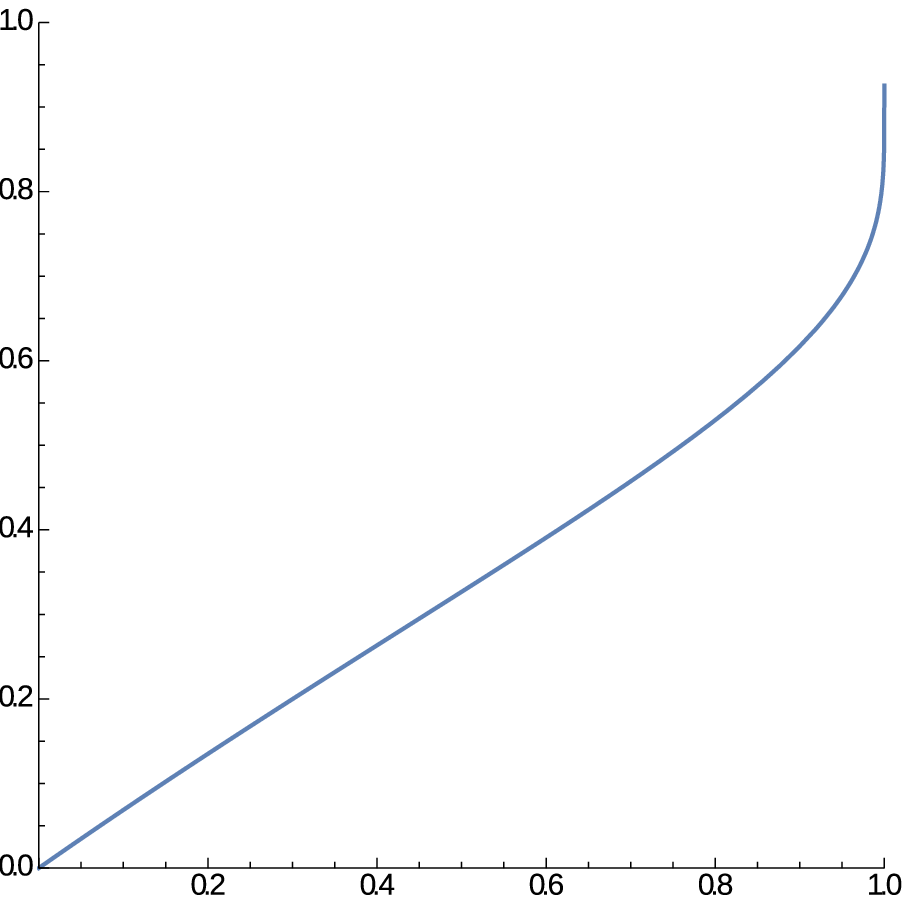,height=1.95in}}
     \hspace{0.3 in}
  \subfigure[$d=0.9$.]{
    	\psfig{file=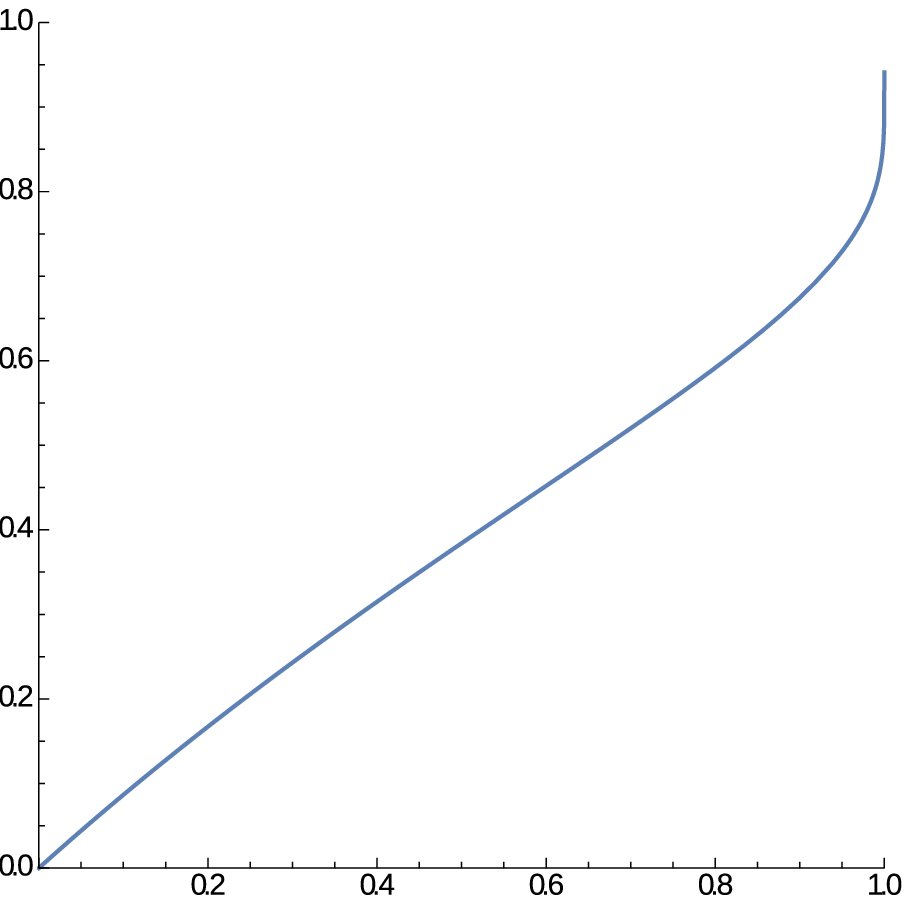,height=1.95in}}
  \subfigure[$d=1$.]{
    	\psfig{file=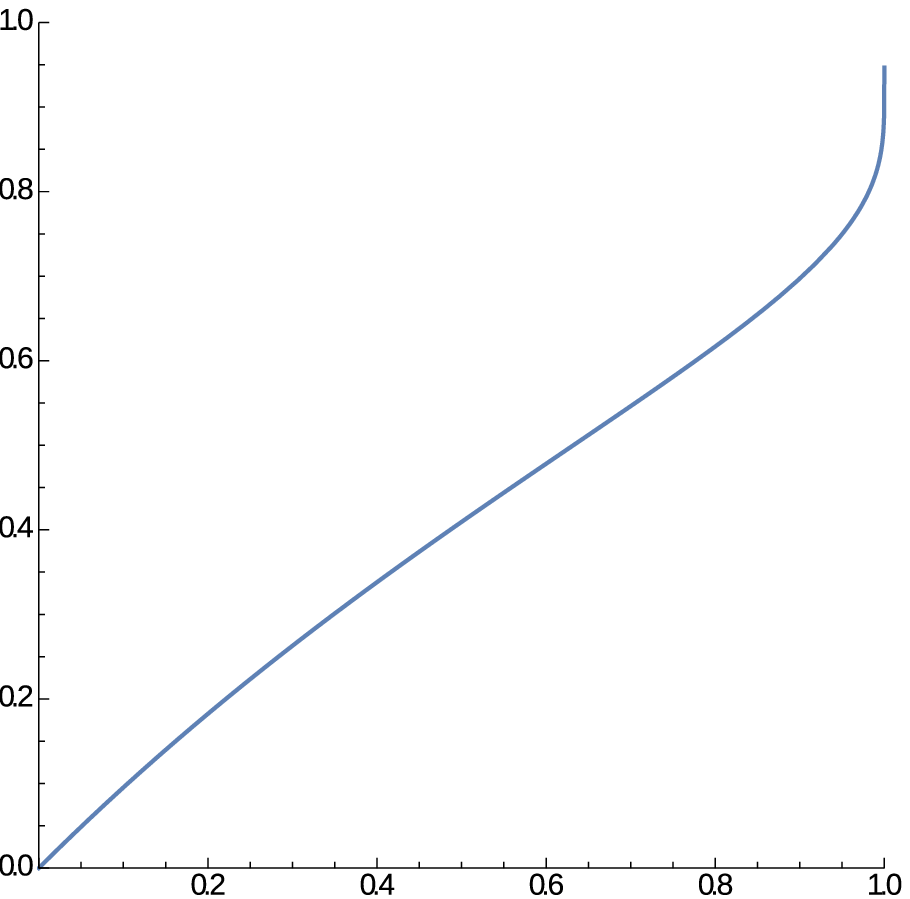,height=1.95in}}
     \hspace{0.3 in}
  \subfigure[$d=2$.]{
    	\psfig{file=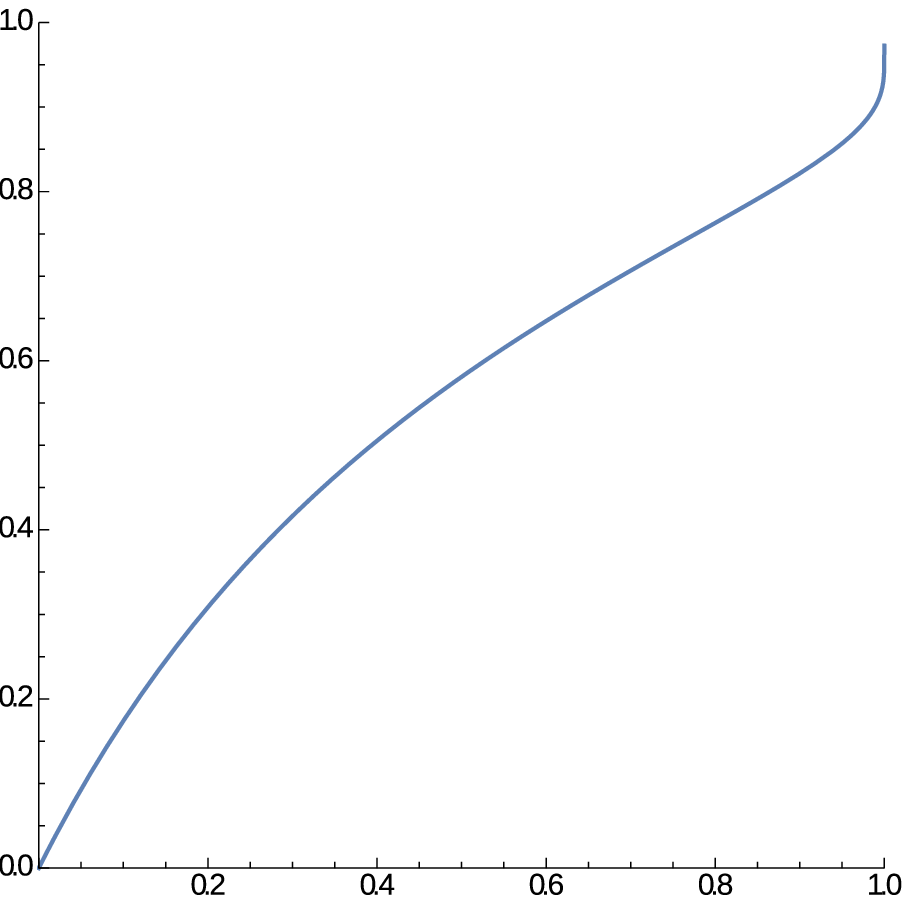,height=1.95in}}
  \subfigure[$d=5$.]{
    	\psfig{file=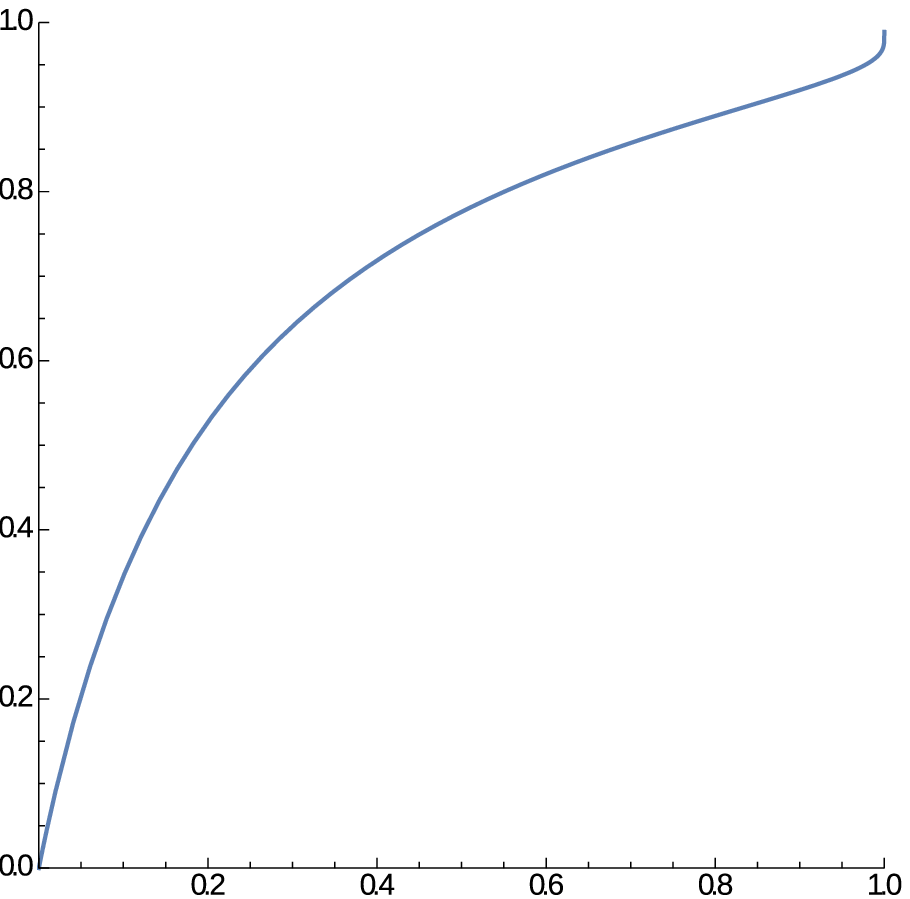,height=1.95in}}
     \hspace{0.3 in}
  \subfigure[$d=10$.]{
    	\psfig{file=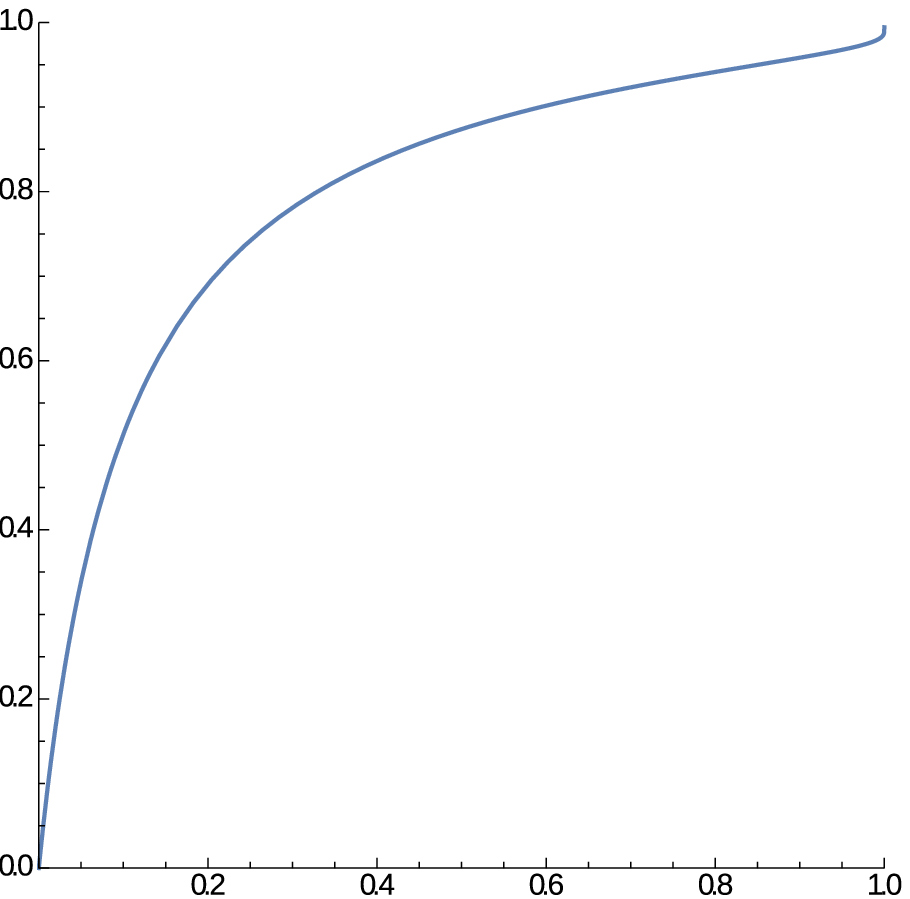,height=1.95in}}
        \caption{Epidemic threshold given by mean field theory $q_0(p,q)$ for 
selected values of $d$.
        }
        \label{fig.1}
\end{figure}

Contrary to the findings of other authors, the data obtained by the simulation 
doesn't seem to show a dependence of the number of infected agents 
at equilibrium, or of the epidemic threshold, exclusively by the mere ratio 
$p/q$, as happens in different, typically continuous-time, models. The epidemic 
threshold $q_0(p,d)$ is only approximately linear for small values of $p$ and 
$d$, which is what we expect to matter, heuristically, if we were to take a 
sort 
of continuum time limit of our model. In fig. \ref{fig.1} we plotted the 
epidemic threshold $q_0(p,q)$ for selected values of $d$. 

\section{Results of the simulation and future work}

\begin{figure}[ht]
        \centering
  \subfigure[$d=0.3$.]{
    	\psfig{file=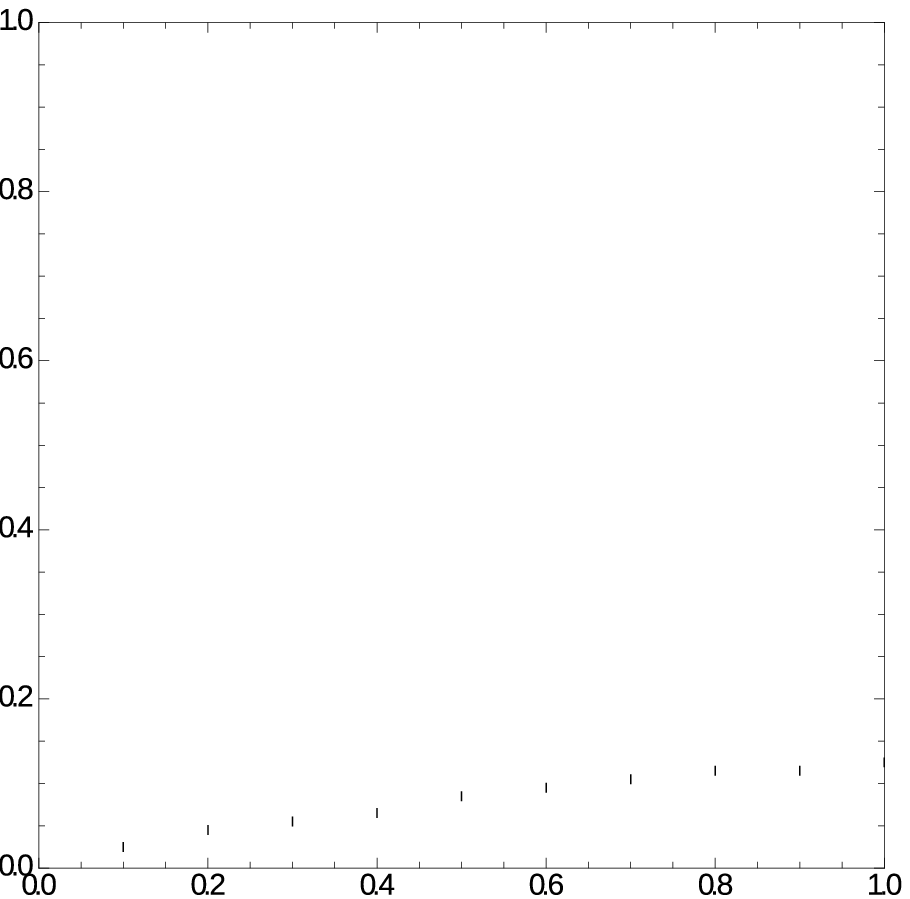,height=1.95in}}
     \hspace{0.3 in}
  \subfigure[$d=0.5$.]{
    	\psfig{file=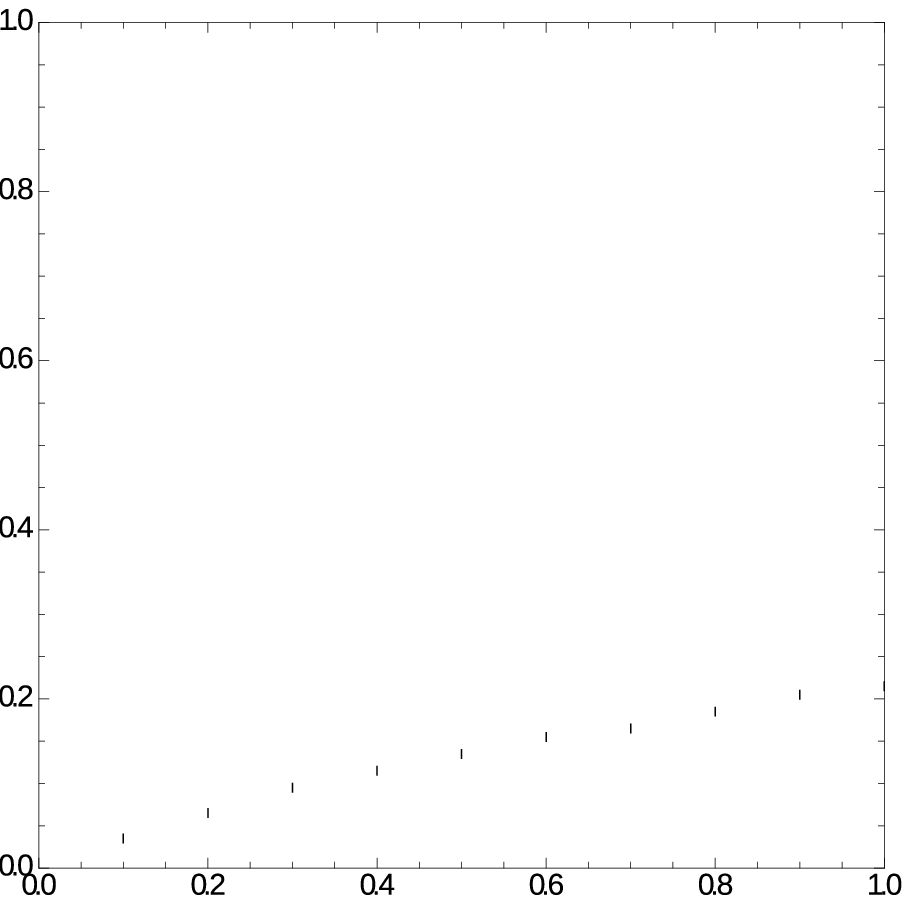,height=1.95in}}
  \subfigure[$d=0.7$.]{
    	\psfig{file=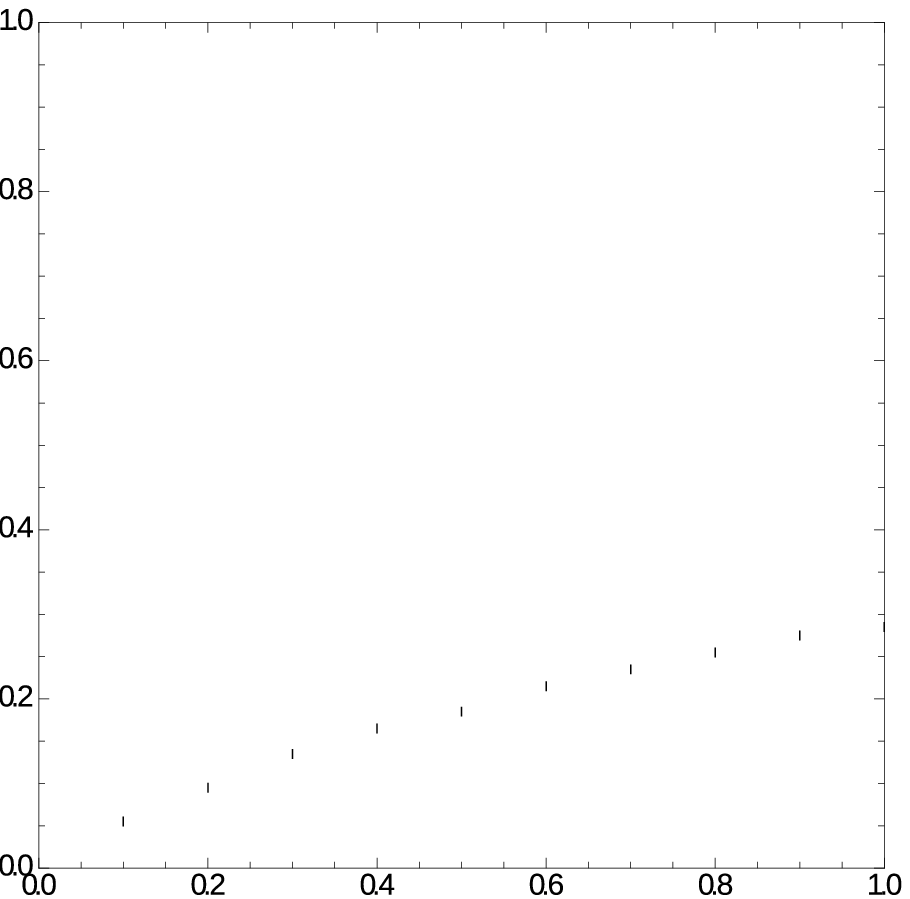,height=1.95in}}
     \hspace{0.3 in}
  \subfigure[$d=0.9$.]{
    	\psfig{file=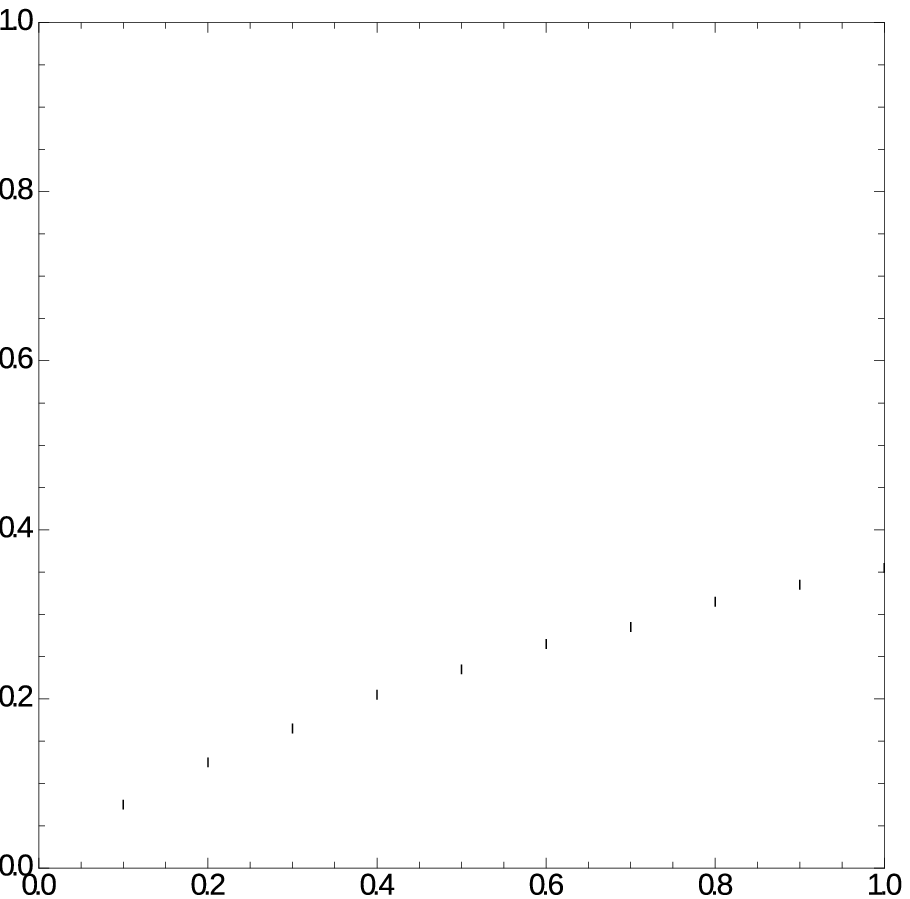,height=1.95in}}
  \subfigure[$d=1$.]{
    	\psfig{file=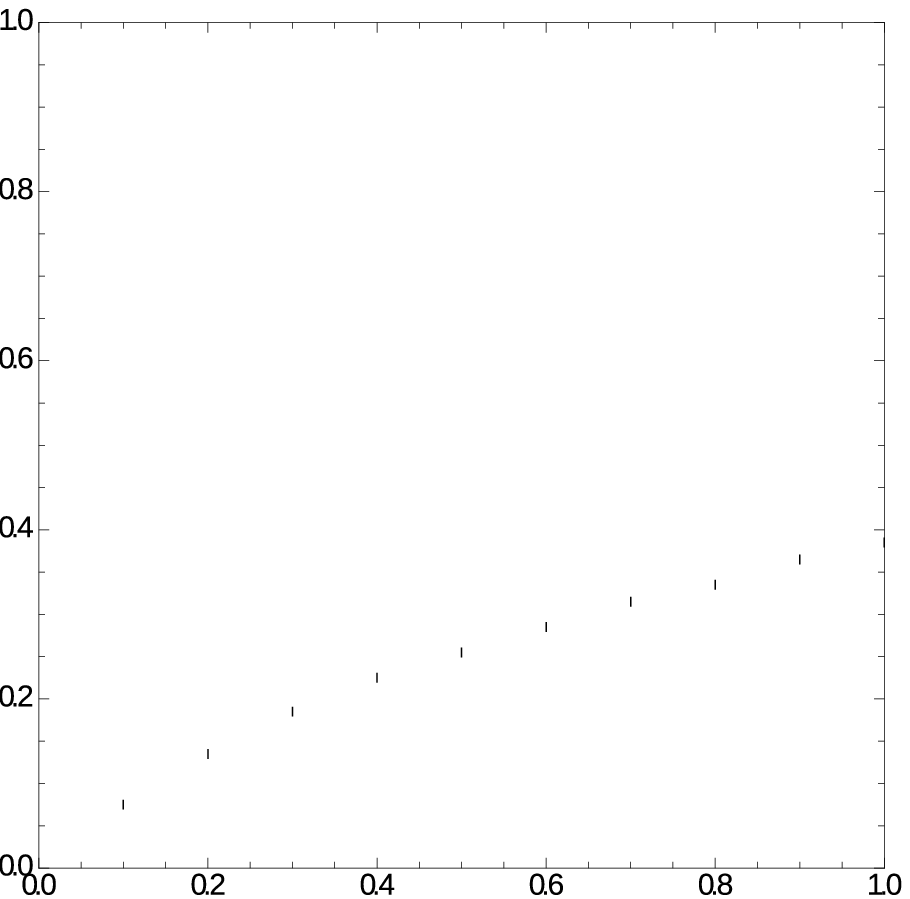,height=1.95in}}
     \hspace{0.3 in}
  \subfigure[$d=2$.]{
    	\psfig{file=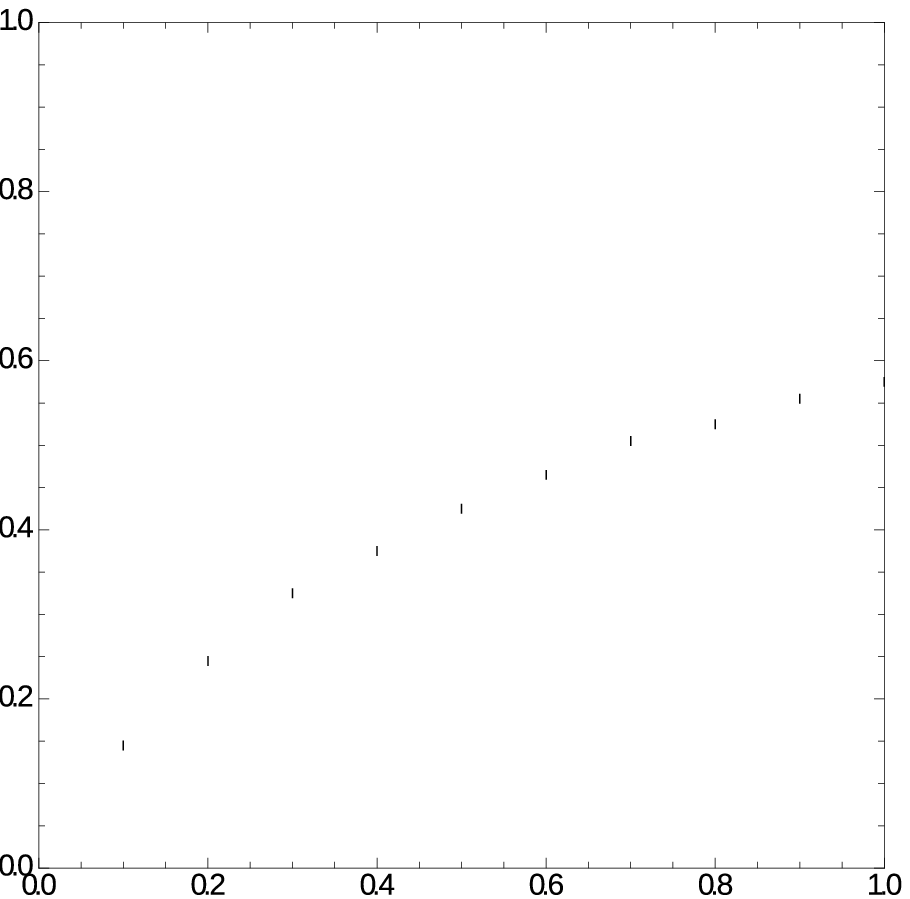,height=1.95in}}
  \subfigure[$d=5$.]{
    	\psfig{file=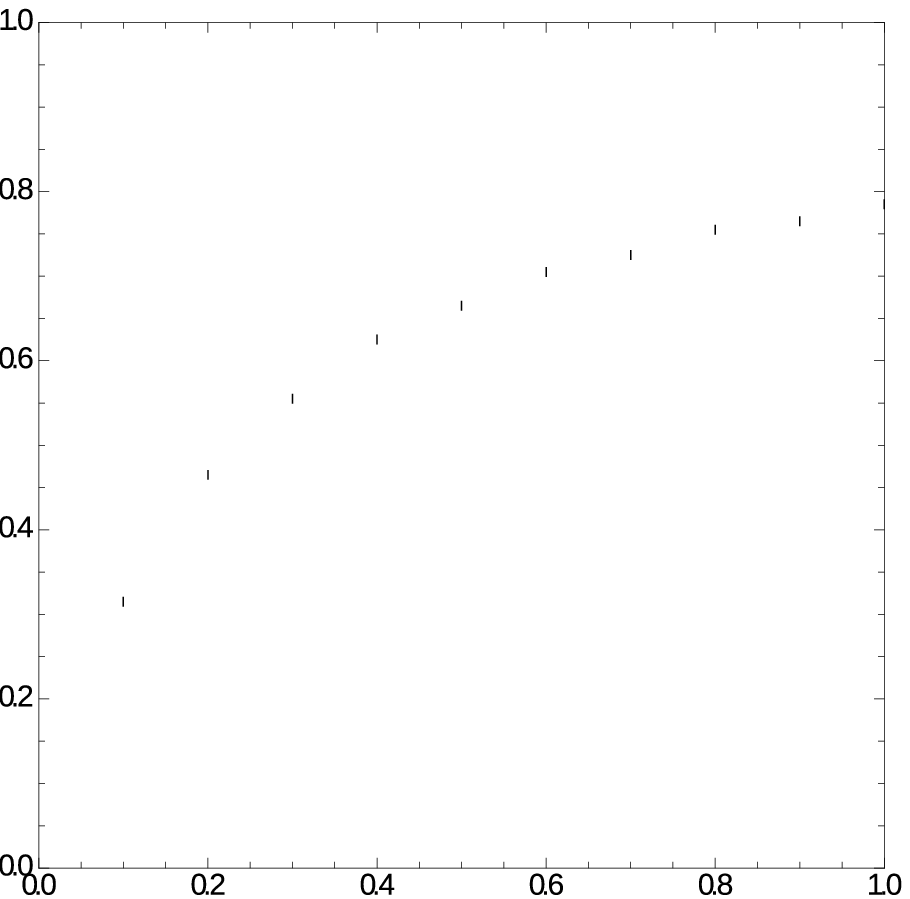,height=1.95in}}
     \hspace{0.3 in}
  \subfigure[$d=10$.]{
    	\psfig{file=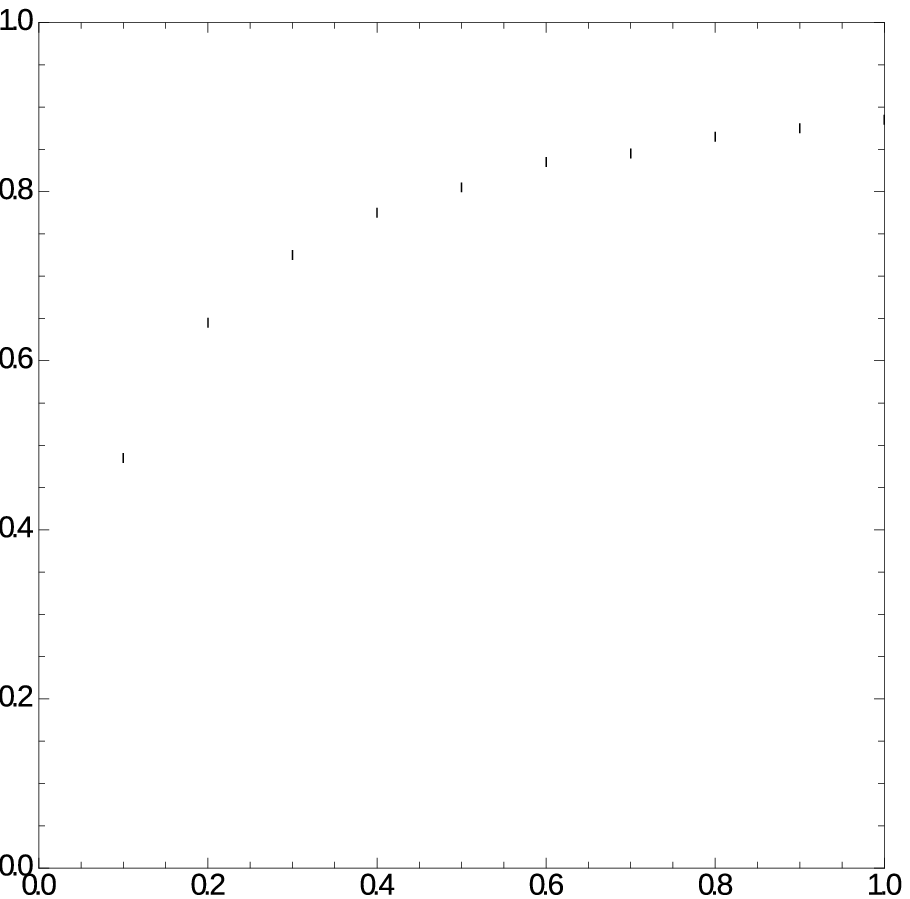,height=1.95in}}
        \caption{Empirical epidemic threshold $q^*(p,q)$ for selected values of 
$d$.
        }
        \label{fig.2}
\end{figure}

The observed fraction of infected agents at equilibrium $f_\infty$ 
depends on all the three parameters: the infection probability $p$, the 
disinfection 
probability $q$ and the density of agents $d$. There appear to be a value 
$q^*(p,d)$ such that if $q > q^*$ then $f_\infty=0$ while if $q < q^*$ then 
$f_\infty \neq 0$, as the mean field theory predicts. $q^*$ is increasing 
both in $p$ and in $d$, as it can be easily expected. $q^*$ 
is again, as predicted by mean field theory, \emph{not linear} in $p$, and so 
there's no ``epidemic threshold'' 
\emph{depending simply on the ratio $p/q$}. In fig. \ref{fig.2} we see some 
plots 
of $q^*(p,d)$ for selected values of $d$. 

The empirical results are qualitatively similar to the predictions of mean 
field theory, but there are some quantitative discrepancies which are stronger 
for small densities of infected agents. We believe that the discrepancies are 
mostly due to the fact that, ultimately, the mean field model is an ergodic 
Markov chain while the real model, which we simulate, is not actually ergodic 
as there is a state (no infected agents at all) which is attracting. Simply 
said, in mean field theory, where we consider only one agent, it can always get 
infected, while in the real model when there are no longer any infected agents 
the infections has no chance to reignite itself. Also, when in the real model 
the density of infected agents is small enough the chance to interact with one 
of the few remaining infected agents is practically negligible and unless $q$ 
is extraordinarily small the infections dies out fast. Note also that in our 
model the probabilities of infection and disinfection $p$ and $q$ are actual 
probabilities of events happening upon intersection of the trajectories of the 
agents, not the infection and disinfection frequencies (which are 
observable random variables and not parameters of the model). 

Concluding, we proposed a model for the propagation of a malware epidemic 
between mobile agents moving randomly on a plane, regular lattice. From 
the purely mathematical point of view changing the dimension of the lattice 
would probably mean a lot, since it would impact the probability of 
intersection of the random walks, but we fail to see a practical application 
for higher dimensional lattices. It would be very interesting anyway to change 
the topology of the environment in which the agent move: for example, the 
agents could be constrained by having malware spreading along a 
graph of connections which is more general than a simple square lattice. This 
would rise the interesting problem of finding an optimal containment strategy
for the epidemic (or even just a better one) by modulating the probabilities of 
infection and disinfection depending on the topological properties of the 
graph. 

\subsection*{Acknowledgments}

We thank the Department of Mathematics of the University of Tor Vergata for 
kindly providing all the computing resources needed for this work.

\end{document}